\documentclass{article}

\usepackage{arxiv}

\usepackage[utf8]{inputenc} 
\usepackage[T1]{fontenc}    
\usepackage{hyperref}       
\usepackage{url}            
\usepackage{booktabs}       
\usepackage{amsfonts}       
\usepackage{nicefrac}       
\usepackage{microtype}      
\usepackage{amsmath}
\usepackage{lipsum}
\usepackage{wrapfig}
\usepackage{graphicx}
\usepackage{newfloat}
\usepackage{listings}
\usepackage{amssymb}
\usepackage{booktabs}
\usepackage{subcaption}
\graphicspath{ {./images/} }

\title{Artificial Intelligence in Open Source Software Engineering: A Foundation for Sustainability}

\author{
  S M Rakib UI Karim \\
  Dept. of Electrical \& Computer Engineering\\
  University of Missouri\\
  Columbia, Missouri, United States \\
  \texttt{skarim@missouri.edu} \\
   \And
  Wenyi Lu \\
  Dept. of Computer Science \\
  University of Missouri\\
  Columbia, Missouri, United States \\
  \texttt{wldh6@mail.missouri.edu} \\
  \And
  Sean Goggins \\
  Dept. of Electrical \& Computer Engineering\\
  University of Missouri\\
  Columbia, Missouri, United States \\
  \texttt{gogginss@missouri.edu} \\
}

\begin{document}
\maketitle
\begin{abstract}
Open-source software (OSS) is foundational to modern digital infrastructure, yet this context for group work continues to struggle to ensure sufficient contributions in many critical cases. This literature review explores how artificial intelligence (AI) is being leveraged to address critical challenges to OSS sustainability, including maintaining contributor engagement, securing funding, ensuring code quality and security, fostering healthy community dynamics, and preventing project abandonment. Synthesizing recent interdisciplinary research, the paper identifies key applications of AI in this domain, including automated bug triaging, system maintenance, contributor onboarding and mentorship, community health analytics, vulnerability detection, and task automation. The review also examines the limitations and ethical concerns that arise from applying AI in OSS contexts, including data availability, bias and fairness, transparency, risks of misuse, and the preservation of human-centered values in collaborative development. By framing AI not as a replacement but as a tool to augment human infrastructure, this study highlights both the promise and pitfalls of AI-driven interventions. It concludes by identifying critical research gaps and proposing future directions at the intersection of AI, sustainability, and OSS, aiming to support more resilient and equitable open-source ecosystems.
\end{abstract}

\section{Introduction}
Open-source software is a total game-changer in today's digital world! It offers publicly available source code that encourages openness, collaboration, and customization. How cool is that? This awesome availability lets individuals and companies dive into the code, tweak it, and share it, all while customizing it to fit their unique needs across different industries \cite{hassri2023impact}.  The principles underpinning open source software, including the essential freedoms to run programs for any purpose, study and modify their source code, redistribute them, and share modified versions to benefit the community form the very foundation of OSS. While the term “Open Source Software” (OSS) is frequently employed, it is essential to acknowledge the broader concept of Free/Libre and Open Source Software (FLOSS), which emphasizes the fundamental rights of users concerning the software \cite{signorini2019open}.

OSS is a key part of supporting important global infrastructure, so developers, organizations, and policy-makers are all very concerned about its long-term viability and ongoing maintenance, which is often called sustainability \cite{dey2024cross}.  In this case, sustainability means that OSS projects can stay alive and useful even when the community and technology change \cite{ye2021open}. The health of thousands of OSS projects is directly related to the stability and security of the digital world as a whole, so it is essential that they stay healthy. The OSS ecosystem, on the other hand, has a hard time staying alive \cite{sun2024sustain}.  There is a well-known problem called the "tragedy of the commons" that points out a basic imbalance: a lot of people benefit from OSS, but only a small number of them actively help keep it up and running and improve it \cite{sun2024sustain}. This dynamic frequently leads to tangible risks such as project abandonment, the proliferation of unaddressed security vulnerabilities, and inadequate end-of-service procedures as projects decline. The growing complexity of modern software systems further exacerbates these sustainability issues, further straining the typically limited resources of OSS projects \cite{cabotsustainability}.

A potential game-changer for making OSS more sustainable, artificial intelligence (AI) has emerged in response to these pressing issues. There are numerous opportunities to increase the overall quantity and quality of software as AI becomes more prevalent in many software development domains \cite{treude2025generative}.  The use of AI-powered tools and intelligent bots to assist with various software development and maintenance tasks is becoming more and more popular, particularly in the open source community \cite{bildirici2024open}.  OSS projects could benefit greatly from AI's ability to automate numerous tasks, analyze large datasets, and provide insightful analysis \cite{zhang2023ai}.

Even though artificial intelligence (AI) is growing in popularity and is known to be helpful in software engineering, a thorough, organized, and current understanding of its specific applications, range of advantages, inherent challenges, and significant ethical implications in the context of the sustainability of open-source software is still lacking. The literature clearly shows a \textbf{synthesizing gap} and a \textbf{recency gap}. Individual studies examine particular AI applications, including particular methods for identifying vulnerabilities.  In order to fully comprehend how new AI technologies interact with the unique social and technical dynamics of the OSS ecosystem, many previous evaluations have either examined AI in software engineering generally or have not gone into sufficient detail. The quick development of advanced AI capabilities, especially Large Language Models (LLMs), widens this gap. This is because their full potential and specific effects on the often distributed, collaborative, and volunteer-driven OSS environment have not yet been fully explored and evaluated in a unified framework.  Our understanding of how the latest developments in AI can responsibly support the long-term viability of open source software projects is thus seriously lacking.  This disparity is widened by the rapid development of sophisticated AI capabilities, particularly Large Language Models (LLMs). Their full potential and specific effects on the often distributed, collaborative, and volunteer-driven OSS environment have not yet been fully explored and evaluated within a single framework, which is the reason for this.  We now know very little about how the latest developments in AI can sustainably contribute to the long-term sustainability of open source software projects.

To bridge this critical knowledge gap, the present work systematically explores the evolving landscape of research that connects artificial intelligence with sustainable practices in open-source software development. The focus will be only on scholarly research papers investigating how AI enables OSS projects to overcome the several obstacles they confront, guaranteeing long-term viability. User’s request dictates that this review exclude blog posts, websites, and other non-academic materials. Specifically, this review aims to:

\begin{itemize}
\item Delineate the core sustainability challenges confronting OSS projects.
\item Establish the foundational role of AI in general software engineering.
\item \textbf{Systematically explore the diverse applications of AI in enhancing OSS sustainability} across critical dimensions, including automated maintenance and bug resolution, community health assessment, new contributor onboarding and mentorship, enhanced security and vulnerability management, and the automation of routine project tasks.
\item \textbf{Critically discuss the inherent challenges, limitations, and ethical implications} associated with leveraging AI in OSS, encompassing data requirements, potential biases, issues of trust and transparency, and the imperative to maintain the crucial human element.
\item \textbf{Identify promising directions for future research}, aiming to guide subsequent scholarly efforts in this increasingly vital interdisciplinary field.
\end{itemize}

This paper seeks to systematically synthesize the current state of Artificial Intelligence (AI) integration within the open-source ecosystem, critically assessing its diverse effects on software development, community health, and long-term sustainability. This review delineates critical opportunities for utilizing AI to bolster the resilience of open-source projects and communities, while concurrently tackling the intrinsic challenges associated with security vulnerabilities, ethical issues, and governance deficiencies. Additionally, it investigates the developing model of symbiotic human-AI collaboration within open-source frameworks, suggesting a strategic vision to enhance this relationship for the advancement of collective intelligence and sustainable innovation. This paper aims to be a fundamental resource for researchers, practitioners, and policy-makers, informing strategic decisions and guiding the creation of robust frameworks that facilitate a resilient and sustainable open-source ecosystem, promoting a synergistic relationship between advanced AI and the collaborative ethos of open source.

\section{Research Methodology}
\begin{wrapfigure}{r}{0.42\textwidth}
  \centering
  \includegraphics[width=0.4\textwidth]{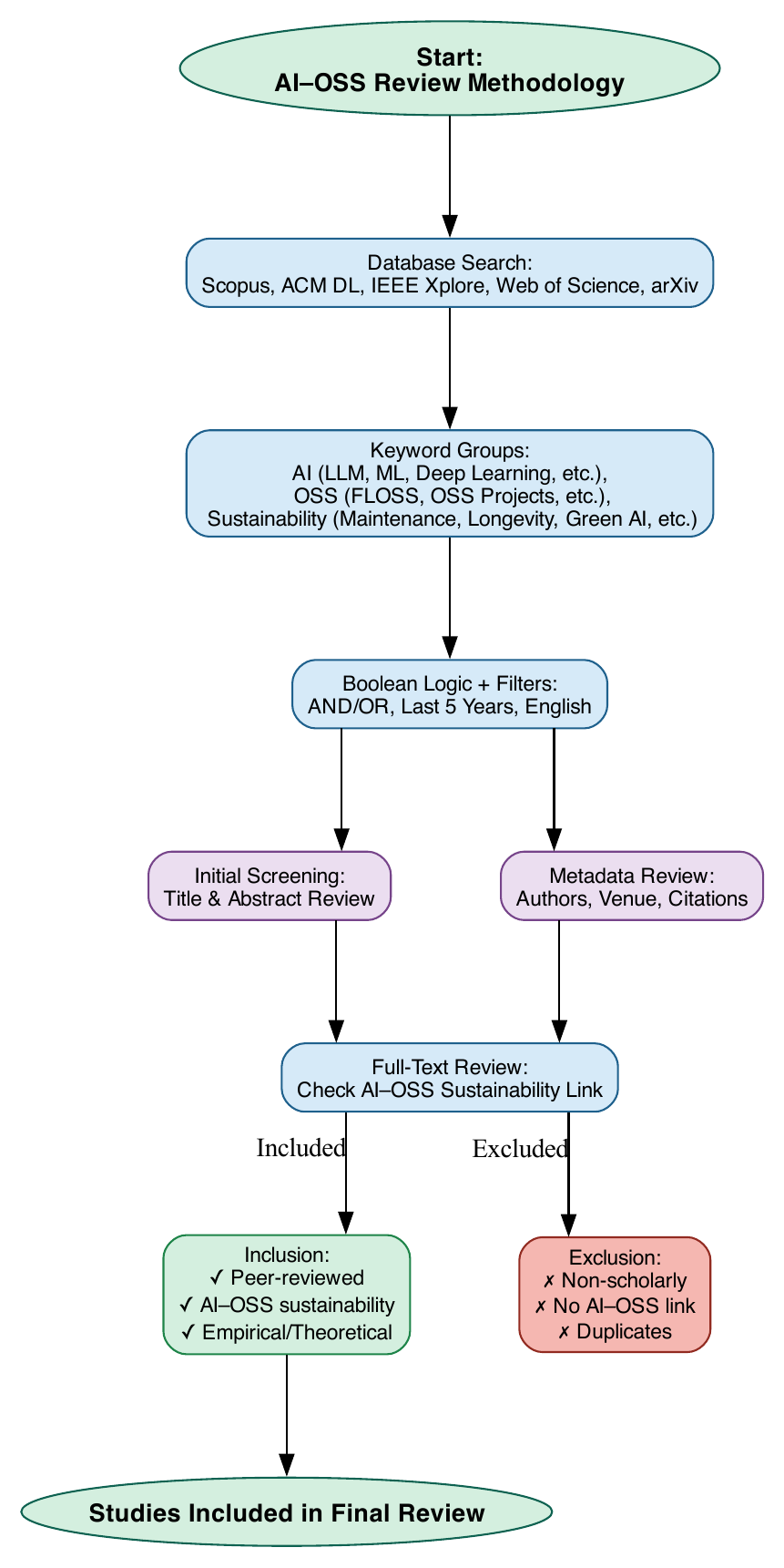}
  \caption{AI-OSS Systematic Review Methodology}
  \label{fig:methodology}
\end{wrapfigure}

This literature review was conducted in a systematic way to make sure that it covered all the important academic literature while also keeping bias to a minimum. The method included a clear plan for finding relevant papers, strict rules for choosing those papers, and a systematic way to put the results together and look at them. A PRISMA flow diagram is a popular and effective tool for visually representing the process of conducting a systematic literature review. Figure \ref{fig:methodology} illustrates that the screened, deemed eligible, and ultimately included in the review are illustrated, providing a clear visual summary of the selection process.

\subsection{Search Strategy}
The primary literature search was conducted using Scopus, Web of Science, ACM Digital Library, IEEE Xplore, and arXiv, among other prominent academic databases and digital libraries. The breadth of coverage in AI, computer science, and software engineering led to the selection of these platforms. We used a combination of keywords to find papers that discussed the crossroads of AI and the long-term viability of open-source software. The main search string included terms associated with AI, OSS, and sustainability, employing Boolean operators (AND, OR) to expand the search parameters while ensuring relevance. The investigation concentrated on publications from the past five years to encompass the latest advancements, considering the swift progression of AI technologies.

\subsection{Paper Selection Criteria}
A multi-stage screening process was employed to select the most relevant papers for this review:

\begin{enumerate}
\item \textbf{Initial Screening (Title and Abstract):} Papers identified through the search queries were first screened based on their titles and abstracts. This stage aimed to quickly filter out irrelevant studies.

\item \textbf{Full-Text Review:} Papers that passed the initial screening underwent a thorough full-text review. This detailed examination ensured that the content directly addressed the application of AI in enhancing OSS sustainability.

\item \textbf{Inclusion Criteria:}
\begin{itemize}
\item Published as scholarly research papers (e.g., journal articles, conference proceedings, peer-reviewed workshop papers).
\item Directly investigated the application of AI techniques to address sustainability challenges in OSS projects.
\item Provided empirical evidence, theoretical frameworks, or comprehensive reviews relevant to the topic.
\item Published in English.
\end{itemize}

\item \textbf{Exclusion Criteria:}

\begin{itemize}
\item Non-academic sources (e.g., blog posts, news articles, whitepapers not peer-reviewed).

\item Papers that discussed AI or OSS in isolation, without a clear connection to OSS sustainability.

\item Duplicate entries.

\item Papers focusing solely on general software engineering without specific implications for OSS.
\end{itemize}
\end{enumerate}

This systematic selection process ensured that the review is based on high-quality, relevant academic literature.

\subsection{Organization of Findings}
Using a thematic synthesis approach, the main ideas and results from the chosen literature were put together. After reading and taking notes on the full texts, we found and sorted the main ideas about the problems with OSS sustainability and AI applications. This process was repeated several times:

\begin{enumerate}
\item \textbf{Identifying Core Challenges:} Extracting and consolidating the primary sustainability challenges faced by OSS projects as discussed in the literature.

\item \textbf{Mapping AI Applications:} Grouping AI techniques and tools based on the specific sustainability challenges they aim to address. This led to the identification of distinct application areas such as automated maintenance, community health analysis, and security management.

\item \textbf{Analyzing Techniques and Implications:} For each application area, a deeper dive into the specific AI algorithms, models, and their practical and ethical implications was conducted.

\item \textbf{Structuring the Review:} The themes that were found naturally made up the structure of the next parts of this paper, which are called "The Landscape of Open-Source Software Sustainability Challenges." “Artificial Intelligence in Software Engineering: A Foundation for Sustainability,” “Applying Artificial Intelligence to Enhance Open-Source Software Sustainability” (which is further broken down into specific AI application subsections), and “Challenges, Limitations, and Ethical Considerations of AI in OSS Sustainability.” This organization makes it easy to go from finding problems to suggesting solutions, talking about their effects, and planning for the future.
\end{enumerate}

\section{The Landscape of Open-Source Software Sustainability Challenges}
Keeping open-source software projects successful and going for a long time is hard work that involves a lot of different problems. One of the biggest problems is making sure that contributors stay involved and keep coming back.   Because they are run by volunteers, many open-source projects have a hard time keeping the contributions of current contributors and getting new ones on a regular basis \cite{sethanandha2010managing}.  It's important to know why people contribute to OSS and what makes them stop contributing in order to keep these communities going \cite{santos2024software}.    Also, for OSS to be healthy in the long run, it is important to welcome new contributors and help them get involved in the project \cite{santos2024software}.   A strong and active user community that thinks they can help the project in any way, such as coding, writing documentation, or anything else, is also important for long-term success \cite{ye2021open}.

Another big problem is that there isn't enough money or resources available \cite{alami2024free}.  Many open source sustainability projects rely on the selfless work of volunteer developers and maintainers, who don't have a lot of money to work with \cite{alami2020sustainability}.  Long-term financial models must be kept up in order for these projects to keep getting funding, new features, and important security updates.   To meet these financial needs, people in the open source community have thought about and looked into different business models.  Some of these methods include setting up nonprofits to run projects, getting licenses to use content for business purposes, letting users pay for access to premium features or support, and creating hybrid models that use more than one source of funding \cite{signorini2019open}.

OSS projects are always having problems with \textbf{code quality and dealing with technical debt}.   It can be hard to make sure that the quality of software is always high because many different developers with different levels of skill and coding styles work on FOSS projects \cite{alami2024free}.  OSS projects can also build up a lot of technical debt over time, which can make it harder to develop and maintain them in the future \cite{venters2018software}.  Strong code review processes help cut down on these problems by making sure that contributions meet the project's standards and that the codebase stays healthy and under control \cite{sethanandha2010managing}.

The ongoing danger of \textbf{security vulnerabilities and the need for efficient patch management} creates major problems for the sustainability of OSS.  The open nature of OSS encourages openness, but it also makes it easier for a wider audience, including bad actors, to find security holes.  So, it is very important to quickly find, fix, and send out security updates \cite{dey2024cross}.  OSS projects are especially at risk for harmful code injections, which makes it even more important to use and follow strict security policies during the development process \cite{cabotsustainability}.  Also, as software systems get more complicated, they often depend on many other OSS libraries, which makes a complicated web of possible security problems that need to be carefully managed \cite{terragni2024future}.

Leibmann et al. \cite{leibmann2025reddit} gained an understanding of collaborative governance and identified several OSS governance challenges by examining governance structures and community management at scale with applications of computational social science methods. They highlighted the overall \textbf{community health and governance} of an OSS project is one of the most important factors for its long-term success. A lively and involved community of users and contributors gives everyone a sense of shared ownership and responsibility for the project's health \cite{alami2024free}.  The governance system of an OSS project, including how decisions are made and who leads it, has a big impact on how long quality will last and where the project is going in general \cite{alami2020sustainability}.  Finding and proactively dealing with "community smells," or different types of organizational and social problems that can happen in software projects, is important for keeping a healthy and productive environment for working together \cite{goggins2021open}.

A lot of OSS projects end up being abandoned. Mature and widely used open source software (OSS) projects can be abandoned if they don't get enough support, if the maintainers get burned out, or if the core contributors have other things to do. Community ownership models or getting organizations to help with important open source software (OSS) projects are both important for keeping software assets safe. Because these problems are all connected, we need full OSS sustainability plans. It's interesting how AI can solve more than one problem at a time. AI-powered tools can make code better, make it safer, help the community work better, and help you find contributors. AI's ability to analyze and automate tasks could be helpful in situations where OSS's openness gets in the way of security and governance.

\section{Artificial Intelligence in Software Engineering: A Foundation for Sustainability}
Artificial Intelligence has affected numerous facets of the software development lifecycle, offering a robust arsenal for addressing persistent issues and enhancing efficiency. From requirements engineering and design to development, testing, release, and maintenance, AI is being increasingly utilized \cite{batarseh2020application}. Advanced AI systems, particularly Large Language Models (LLMs), are boosting software development productivity and promising a future where humans and AI work together \cite{terragni2024future}. Growing research focuses on defining a structured taxonomy of software engineering tasks where AI can provide meaningful assistance, beyond code generation \cite{gu2025challenges}.

In the field of \textbf{code analysis and quality assurance}, AI is substantially making strides. A number of code quality issues, such as code smells, possible bugs, and security vulnerabilities, can be detected by AI-powered tools that employ static and dynamic analysis techniques \footnote{https://www.ibm.com/think/insights/ai-code-review; Accessed: 2025-05-05}. Machine learning algorithms are used to understand the underlying semantics of code and identify typical anti-patterns that may cause issues.  Moreover, artificial intelligence can analyze current codebases and provide recommendations for targeted changes aimed at enhancing code efficiency, readability, and long-term maintainability. In software engineering, artificial intelligence is also changing the procedures of testing and validation.  AI-driven technologies can create test cases, evaluate the degree of code coverage achieved through testing activities, and identify potential failure points that might otherwise go unnoticed \cite{zhang2023ai}. AI enhances the efficiency and reliability of software testing.  Additionally, AI methods are being investigated to solve flaky tests, which produce inconsistent results, improving the testing suite's reliability.

Another main area of influence is the possibility of \textbf{AI-driven automation in development and maintenance}. AI engineering enables the automation of many tedious and repetitive tasks related to software development, including testing, debugging, and continuous system maintenance \cite{budiardjo2024roadmap}. Additionally, AI is increasingly automating code generation and suggesting intelligent code completion to developers.  AI algorithms can also analyze code and suggest optimal refactoring strategies to improve software quality and performance \cite{zhang2023ai}.

AI is gaining traction in \textbf{software project management}. AI applications are being explored for tasks such as project planning, accurately estimating the effort required for development tasks, and proactively managing potential project risks \cite{crawford2023ai}.  By leveraging AI techniques, software project managers can gain valuable insights from project data, enabling them to make more informed decisions regarding resource allocation, scheduling, and overall project execution \cite{romero2023optimising}. 

The software development lifecycle's widespread applicability of artificial intelligence offers a strong basis for its possible use in tackling the several issues connected with OSS sustainability. AI's potential extends well beyond merely automating code-related tasks; it also encompasses fields vital to OSS health, including community management, security improvement, and project governance. The changing view of artificial intelligence as a cooperative "smart workmate" indicates a closer integration into development processes, therefore implying a changing possibility for the future management and sustainability of OSS projects.

\section{Applying Artificial Intelligence to Enhance Open-Source Software Sustainability}

The strategic use of several artificial intelligence technologies could help solve the problems that open-source software sustainability faces. This section explores specific ways in which AI is being leveraged to enhance the long-term viability and health of OSS projects. Table~\ref{tab:ai_applications_oss} provides a concise overview of the key sustainability challenges faced by open-source software projects and the corresponding Artificial Intelligence techniques or applications that are being explored to address them. It also lists the snippet IDs of the research papers that discuss these applications and highlights the potential benefits of using AI in each area.

\begin{table*}[h]
\centering
\caption{AI Applications for OSS Sustainability}
\label{tab:ai_applications_oss}
\resizebox{\textwidth}{!}{%
\begin{tabular}{|p{4cm}|p{5cm}|p{2.5cm}|p{5cm}|}
\hline
\textbf{Sustainability Challenge} & \textbf{AI Technique/Application} & \textbf{Relevant Snippet IDs} & \textbf{Potential Benefits} \\
\hline
Bug Fixing & LLMs for bug fixing, Neural Machine Translation for patches & \cite{oyeniran2023ai, alsaedi2024leveraging} & Reduced maintenance burden, Improved software stability \\
\hline
Community Health Analysis & Machine Learning for metrics, Sentiment Analysis & \cite{goggins2021open} & Enhanced understanding of community dynamics, Proactive issue identification \\
\hline
Contributor Onboarding & AI-powered chatbots, Recommendation Systems for tasks and mentors & \cite{tan2025revolutionizing, he2023open} & Faster integration of new contributors, Increased participation \\
\hline
Security Vulnerability Detection & AI-driven Static and Dynamic Analysis, NLP for code analysis & \cite{rajapaksha2023enhancing, rajapaksha2022ai} & Enhanced security posture, Proactive identification of weaknesses \\
\hline
Project Management & Bots for task automation, AI for effort estimation and risk management & \cite{cabotsustainability, crawford2023ai} & Increased efficiency, Better resource allocation \\
\hline
Environmental Impact & Energy-efficient AI models, AI-driven co-design for hardware and software optimization & \cite{xie2025data, broekema2024supercode} & Lower energy consumption, Reduced carbon footprint \\
\hline
\end{tabular}%
}
\end{table*}


\subsection{Automated maintenance and bug fixing}
The application of \textbf{AI-powered tools for automated maintenance and bug fixing} is highly promising as these tools enable proactive bug detection, automated patch generation, and intelligent code review. These capabilities significantly reduce maintenance burdens and enhance the overall stability and quality of open-source software projects. This includes several important elements.  Using machine learning techniques to examine code for abnormalities and possible faults, \textbf{Intelligent Bug Detection and Prediction provides} a proactive way to find problems \cite{chilkotiai}. AI can learn from the huge amounts of past bug data in OSS projects, which makes it possible to construct models that can find possible flaws early in the development cycle \cite{zhang2023ai}. AI may also provide developers with real-time feedback on potential flaws as they create code, allowing them to fix them immediately \cite{chilkotiai}. Applying these AI-powered bug detection approaches to open-source projects is the main focus of research, with the goal of improving software stability and reliability. \textbf{Automated Patch Generation and Code Repair} builds on issue detection by looking into how AI and machine learning could be used to automatically create fixes for bugs that have already been found, using fixes that have been learnt from previous bug fixes.  Researchers are exploring Neural Machine Translation methods to train models on broken code and the fixes that can be obtained from open-source software (OSS) repositories. This will let the AI “translate” buggy code into its fixed version \cite{weisz2021perfection}. The rise of Large Language Models (LLMs) has opened up new areas of research in automated bug fixing, demonstrating that they can fix various types of bugs with greater accuracy. Experimental investigations are also assessing the ability of LLM-based agents to automatically patch flaws in open-source software \cite{alsaedi2024leveraging}. These efforts are supported by \textbf{AI-Assisted Code Review and Refactoring}. AI code review could improve OSS code review efficiency and uniformity. AI tools can examine code in real-time, providing open-source project maintainers and contributors with actionable information about potential faults and improvements. These tools can also suggest enhancements to code efficiency and maintainability, contributing to a healthier codebase \cite{zhang2023ai}. The power of LLMs is also being examined for creating intelligent code reviews, potentially speeding the process and allowing human reviewers to focus on more complicated areas \cite{cihan2025automated}. While LLM-based code review tools show promise in enhancing bug detection and promoting best practices, they can also lead to longer pull request closure times and may sometimes provide faulty or irrelevant comments \cite{lin2025open}.

\subsection{Community Health and Activity Analysis}
AI is also proving to be a valuable asset in \textbf{Community Health and Activity Analysis}. The sustainability of an OSS project is heavily reliant on the health and engagement of its community \cite{weld2024making}, and AI can provide powerful tools for assessment. \textbf{Metrics and Models for Assessing OSS Community Health using AI} are being developed and utilized to gain a deeper understanding of project vitality \cite{alsaedi2024leveraging}. AI and machine learning are capable of looking at the huge quantities of trace data from OSS repositories, issue trackers, and communication platforms to give us numbers that tell us how healthy a project is \cite{goggins2021open, goggins2013group}. Tools like YOSHI and csDetector leverage these techniques to map project activity to established community patterns and even identify potential “community smells” that might indicate underlying issues \cite{almarimi2021csdetector}. By analyzing contributor activity patterns using AI, projects can potentially identify early warning signs of contributors becoming less active. Predictive modeling can then be employed to forecast potential attrition, allowing project maintainers to proactively reach out and address any underlying issues. \textbf{AI-Driven Insights for Community Engagement and Growth} can also help OSS projects learn more about the people that contribute to them. AI can look at what contributors want and why they want it, which makes it possible to create more focused and effective ways to get them involved. AI can assist find important people and possible future leaders by looking at how people in the community talk to each other. AI-powered sentiment analysis of community talks can also give you useful information about how healthy the community is generally and point out possible areas of conflict or worry.

\subsection{New Contributor Onboarding and Mentorship}
It is very important for the OSS community to be able to successfully integrate newcomers in order to stay around for a long time. For instance, Zhou et al. \cite{zhou2024emoji} mentioned in their study that communications happening within OSS collaboration platforms, like GitHub, can significantly affect sustainable participation, especially for new contributors, through directly examining contributor engagement and project management. AI is becoming more and more significant in New Contributor Onboarding and Mentorship. Using AI to create personalized onboarding experiences can make it easier for new people to get started. AI mentors built right into OSS platforms can help new contributors in real time while they work on the project \cite{tan2025revolutionizing}. AI can also tailor onboarding materials and recommend specific resources based on a newcomer's self-identified skills and areas of interest. AI-Powered Recommendation Systems for Tasks and Mentors can further streamline the onboarding  \cite{song2024impact}. AI can analyze the project's issue backlog and recommend suitable starting tasks based on a newcomer's indicated skill level. Moreover, AI may evaluate the proficiency and availability of current community members and autonomously suggest suitable mentors for new contributors according to their needs and the project's specifications \cite{he2023open}.

\textbf{Automated Support and Guidance for Newcomers} can aid immediately and relieve busy maintainers. AI-powered chatbots and virtual assistants may answer common inquiries, link to relevant content, and help newbies through contribution workflows. AI can also automatically review new code contributors' contributions to assist them learn and follow the project's coding standards and guidelines. Research reveals that newbies expect AI mentor support in finding an engaging topic to contribute to.

\subsection{Enhanced Security and Vulnerability Management}
The most important thing is to keep OSS safe and sound, and AI is being used to improve security and vulnerability management. AI-Driven Static and Dynamic Analysis for Vulnerability Detection is a powerful tool in this area. AI and ML can look at source code like it's a piece of text and use Natural Language Processing to find possible security holes. AI-powered static and dynamic analysis tools can automatically flag a wide range of potential bugs and security weaknesses in the code \cite{rajapaksha2023enhancing}. These models can achieve high accuracy in detecting vulnerabilities in code \cite{cabotsustainability}.

Beyond detection, AI can also contribute to \textbf{Predictive Modeling of Security Risks}. By analyzing historical vulnerability data from OSS projects, AI can potentially identify patterns and predict future security risks. Machine learning models can be trained to recognize subtle anomalies and patterns in code that might indicate underlying security weaknesses that could be exploited.

Furthermore, AI can facilitate Automated Response and Mitigation Strategies \cite{keller2024ai}. AI technologies can provide targeted remedies for vulnerabilities detected in open-source software code, potentially expediting the patching procedure. The capability of AI to automate the generation of fixes for identified security vulnerabilities is a current research focus, offering expedited responses to significant security concerns \cite{zhang2024fixing}.

\subsection{AI-Powered Bots and Agents for Project Management}
Managing OSS projects well frequently means doing a lot of the same things over and over again. AI-Powered Bots and Agents are becoming useful tools for this.  Automating everyday tasks and administrative work can provide maintainers more time to work on the most important parts of development.  You can employ bots to take care of boring but important chores, such as automatically giving labels to new problems that are reported \cite{cabotsustainability}. AI-powered bots are also being made to do more complicated things, such as automatically sorting and managing issues in OSS project tracking systems \cite{jahanshai2022ai}.

\begin{table*}[h]
\centering
\caption{Comparative Analysis of AI Techniques for Key OSS Sustainability Challenges}
\label{tab:comparative_analysis}
\resizebox{\textwidth}{!}{%
\begin{tabular}{|p{2.5cm}|p{2.5cm}|p{3.5cm}|p{3cm}|p{4.5cm}|p{4.5cm}|}
\hline
\textbf{Sustainability Challenge} & \textbf{AI Technique Category} & \textbf{Specific AI Techniques} & \textbf{Key Research Papers (Snippet IDs)} & \textbf{Strengths} & \textbf{Limitations/Challenges} \\
\hline
Automated Bug Fixing & Large Language Models (LLMs) & Code generation for patches, Defect prediction & \cite{ye2021open} & High success rates in fixing various bug types, Automated patch generation, Requires only buggy code & Potential for overfitting, Accuracy depends on training data quality \\
\hline
Community Health Analysis & Machine Learning & Metric analysis, Sentiment analysis, Contributor activity & \cite{goggins2021open, colt2023introduction} & Provides quantitative insights into community dynamics, Identifies potential issues & Relies on availability and quality of trace data, Interpretation can be complex \\
\hline
Contributor Onboarding & Recommendation Systems & Task recommendations, Mentor recommendations, Chatbots & \cite{tan2025revolutionizing, he2023open} & Personalizes onboarding, Reduces burden on maintainers, Facilitates quicker integration & Effectiveness depends on understanding contributor skills and project needs \\
\hline
Vulnerability Detection & Static/Dynamic Analysis \& ML & Code scanning, Pattern recognition, Predictive modeling & \cite{batarseh2020application, weber2023sustainability} & Proactive identification of security weaknesses, Can analyze large codebases efficiently & May produce false positives, Requires continuous updating with new vulnerability data \\
\hline
Code Review & Large Language Models (LLMs) & Automated feedback, Suggesting improvements & \cite{cihan2025automated, lin2025open} & Provides consistent and timely feedback, Can identify common issues & May lack deep contextual understanding, Risk of generating noisy or irrelevant comments \\
\hline
\end{tabular}%
}
\end{table*}

These intelligent agents can provide \textbf{Intelligent Assistance for Maintainers and Contributors}. Smart bots can learn from an OSS project's massive data set to become talented assistants that can intelligently triage bug reports and ensure they contain enough information. Bots can also assess developer activity and suggest candidates for project tasks\cite{cabotsustainability}.

Finally, AI can help people in the community talk to and work with one other. Chatbots that employ AI can help with community assistance by addressing common inquiries and pointing users to useful sites. Bots can also look at the content of problems and pull requests to identify contributors or reviewers who might have the right skills to give useful comments.

\subsection{Green AI and Sustainable Practices in OSS Development}
As AI becomes more common in software development, it's important to think about how it affects the environment \cite{hidalgo2023sustainable, danushi2024environmentally}. Sustainable practices in OSS development and green AI are becoming more and more significant \cite{calero2024addressing}. Looking into the Environmental Impact of AI in Software Development shows that training and using complicated AI models can use a lot of energy, which means more electricity is needed and a big carbon footprint \cite{xie2025data, kumar2024balancing}. Therefore, the need for sustainable AI development practices to minimize these environmental consequences is crucial \cite{cruz2024innovating}. In particular, there has been increasing concern surrounding the life-cycle energy costs of models covering training, inference, retraining, and deployment across distributed OSS environments.

Researchers are exploring Strategies for Developing Energy-Efficient AI Models for OSS, including the use of specialized energy-efficient hardware and the optimization of AI algorithms to reduce their computational complexity \cite{xie2025data, chen2023survey}. Another viable way to cut down on AI's overall energy use is to create machine learning models that don't need as many parameters \cite{wu2024beyond}. Parameter-efficient training strategies such as low-rank adaptation (LoRA) and pruning are also being studied to achieve comparable performance while dramatically reducing the number of trainable weights and compute requirements. SuperCode and other AI-driven co-design approaches are also being investigated as potential means of producing extremely efficient code for future computer hardware, with sustainability being specifically identified as a critical performance measure \cite{batarseh2020application}. These techniques align with broader industry trends towards carbon-aware computing, in which compilation and inference pipelines can dynamically optimize energy consumption based on carbon intensity signals from the power grid.

Furthermore, there is a substantial role for open source in promoting sustainable AI. Better openness, accountability, and community-driven attempts to build AI solutions that are both ethical and ecologically sustainable can result from AI work being open-source. One way that open-source principles can help create a more responsible and sustainable AI ecosystem is by allowing a broad community to review, test, and improve AI algorithms and the data used to train them. This way, biases and security vulnerabilities can be exposed and addressed \cite{bildirici2024open}. In addition, open-source ecosystems enable shared infrastructure (e.g., model checkpoints, datasets, evaluation tools, and reproducibility frameworks), which reduces duplicated computational effort across institutions and lowers the aggregate carbon footprint associated with experimentation and benchmarking.

Several emerging research efforts have also started to incorporate formal sustainability metrics into OSS-AI workflows. For example, carbon accounting frameworks now allow AI developers to estimate and report emissions for model training, enabling transparent environmental assessments. Some initiatives additionally advocate for model cards and dataset documentation that include resource utilization and energy cost disclosures. Such practices could eventually evolve into standardized sustainability reporting conventions for OSS projects that employ AI, complementing current discussions on responsible and trustworthy AI.

\begin{table*}[t]
\centering
\caption{Categorization of Research Papers by AI Technique and OSS Sustainability Focus}
\label{tab:categorization}
\resizebox{\textwidth}{!}{%
\begin{tabular}{|p{5.5cm}|p{5cm}|p{6cm}|}
\hline
\textbf{AI Technique Category} & \textbf{OSS Sustainability Focus} & \textbf{Example Research Papers (Snippet IDs)} \\
\hline
Large Language Models (LLMs) & Automated Bug Fixing & \cite{ye2021open} \\
\hline
Large Language Models (LLMs) & Code Review &  \cite{cihan2025automated, lin2025open} \\
\hline
Machine Learning & Community Health Analysis &  \cite{goggins2021open} \\
\hline
Machine Learning & Vulnerability Detection & \cite{batarseh2020application} \\
\hline
Recommendation Systems & Contributor Onboarding &  \cite{tan2025revolutionizing, he2023open} \\
\hline
Natural Language Processing (NLP) & Code Analysis \& Documentation & \cite{rajapaksha2023enhancing, rajapaksha2022ai} \\
\hline
AI-powered Bots/Agents & Project Management \& Automation &  \cite{weber2023sustainability, crawford2023ai} \\
\hline
Green AI & Energy Efficiency \& Sustainability &  \cite{batarseh2020application, xie2025data, broekema2024supercode} \\
\hline
\end{tabular}%
}
\end{table*}

Building on the overview of AI applications in Table~\ref{tab:ai_applications_oss}, the comparative analysis in Table~\ref{tab:comparative_analysis} summarizes different AI techniques used for tackling specific sustainability challenges in open-source software. It categorizes the AI techniques, lists specific examples, and points to relevant research papers, while also outlining the strengths and limitations associated with each approach. To further illustrate the research landscape, Table~\ref{tab:categorization} categorizes the research papers included in this literature review based on the primary AI technique employed and the specific aspect of open-source software sustainability that they address. This provides a structured view of how different AI methods are applied across various sustainability areas.


In the final analysis, Table~\ref{tab:key_papers} provides an overview of some of the most important research papers in the field. It includes necessary information such as the title, authors, publication year, the primary artificial intelligence technique that was utilized, the emphasis area within open source software sustainability, and a quick explanation of their most important findings. A fast comprehension of the key contributions made to the discipline is made possible by this table.

\section{Challenges, Limitations, and Ethical Considerations of AI in OSS Sustainability}
The use of AI presents considerable potential for improving the sustainability of open-source software; nonetheless, numerous problems, limits, and ethical considerations must be meticulously addressed to guarantee responsible and effective application.

One of the biggest problems is getting the right data for training AI models.  AI models, especially those that use deep learning, usually need a lot of high-quality data to train well and give correct answers \cite{wu2022sustainable}. However, the availability and quality of data from OSS projects can vary considerably \cite{alami2024free}. It is important to deal with problems like data heterogeneity, where datasets from different projects may have different forms and formats, and possible biases that may be present in these datasets, in order to create strong and trustworthy AI solutions for OSS \cite{lin2025open}. 

Another important ethical issue is \textbf{Bias and fairness in AI applications for OSS}. AI algorithms could accidentally keep and even make worse any biases that are already in the data they are trained on \cite{zhang2023ai}. Ensuring fairness and diversity in AI-driven procedures within the various and frequently worldwide groups contributing to OSS is essential. When implementing AI in open source software (OSS) contexts, developers and maintainers need to think about the ethical implications and work to make tools that are fair to all members of the community \cite{budiardjo2024roadmap}.

Building \textbf{trust and transparency in AI-driven processes within OSS communities} is essential for their widespread adoption \cite{alami2020sustainability}. OSS thrives on open collaboration and the trust that contributors place in the project's processes and maintainers \cite{alami2024free}. Consequently, ensuring the transparency and elucidation of AI decision-making is imperative in this scenario. Developers need to understand how AI tools make their suggestions and be sure that they are reliable. It's crucial to think about how to avoid putting too much faith in AI proposals that might not be right and how to have a balanced view of what AI can and can't do in software engineeringneering \cite{lo2023trustworthy}.

The \textbf{potential for misuse and security risks associated with AI tools} within the open-source ecosystem cannot be overlooked \footnote{https://lfaidata.foundation/blog/2024/03/04/open-source-ai-opportunities-and-challenges/}. As AI models and tools become more sophisticated, the risk of malicious actors leveraging them for harmful purposes increases. Addressing security vulnerabilities that may be inherent in AI-based software engineering tools and ensuring their resilience against various forms of attacks are critical for maintaining the integrity of OSS projects \cite{lo2023trustworthy}. Also, there needs to be clear rules and considerable thought given to the issues of intellectual property and the licensing of code made by AI inside the OSS framework.

\begin{table*}
\centering
\caption{Key Research Papers on AI for OSS Sustainability}
\label{tab:key_papers}
\resizebox{\textwidth}{!}{%
\begin{tabular}{|p{6cm}|p{3cm}|p{1cm}|p{3cm}|p{4cm}|p{6cm}|}
\hline
\textbf{Title} & \textbf{Authors} & \textbf{Year} & \textbf{Focus AI Technique} & \textbf{Focus OSS Sustainability Aspect} & \textbf{Key Findings} \\
\hline
SuperCode: Sustainability PER AI-driven CO-DEsign & P. Chris Broekema, Rob V. van Nieuwpoort & 2024 & Large Language Models & Energy Efficiency & Proposes AI-driven co-design for generating efficient code for emerging computing hardware with sustainability as a key metric. \\
\hline
Enhancing Security Assurance in Software Development: AI-Based Vulnerable Code Detection with Static Analysis & SAMPATH RAJAPAKSHA R WASALA MUDIYANSELAGE POLWATTE GEDARA et al. & 2023 & Machine Learning, Explainable AI & Security Vulnerability Detection & Demonstrates high accuracy in detecting vulnerabilities in C/C++ code using ML and provides explanations for developers. \\
\hline
Leveraging Large Language Models for Automated Bug Fixing & Syed Muhammad Ali, Muhammad Asif, Muhammad Usama & 2024 & Large Language Models & Automated Bug Fixing & Shows the potential of LLMs to achieve high success rates in fixing bugs in Java code, outperforming existing APR models. \\
\hline
Revolutionizing Newcomers' Onboarding Process in OSS Communities: The Future AI Mentor & Asif Kamal Turzo, Sayma Sultana, Amiangshu Bosu & 2025 & Design Fiction, AI Mentor Prototype & Contributor Onboarding & Explores the potential of AI as a comprehensive mentor for new OSS contributors, identifying key design strategies. \\
\hline
How to Sustain a Scientific Open-Source Software Ecosystem: Learning from the Astropy Project & Jiayi Sun et al. & 2024 & Mixed Methods (Interviews, Surveys, Analysis) & Overall Sustainability & Examines challenges and opportunities for sustaining scientific OSS, proposing concrete suggestions. \\
\hline
\end{tabular}%
}
\end{table*}

The successful integration of AI into OSS sustainability efforts also depends on overcoming integration challenges and recognizing the crucial human element in OSS development \cite{terragni2024future, yang2023automated}. AI technologies must be developed to effortlessly fit into the existing workflows and practices of OSS projects without creating undue disruption.  It is essential to combine the advantages of automation provided by AI with the crucial contributions of human expertise, creativity, and cooperation, which are important to the success of OSS initiatives. The social dynamics and collaborative ethos that define OSS communities are essential to their longevity, and the integration of AI should seek to augment, rather than obstruct, these critical elements \cite{alami2024free}.


\section{Conclusion and Future Research Directions}
This literature review has examined the rapidly evolving intersection of artificial intelligence and open-source software sustainability, with a focus on how AI-enabled techniques can mitigate the structural, social, and technical challenges that threaten the long-term viability of OSS projects. The analysis shows that researchers and practitioners are increasingly recognizing AI not merely as a productivity aid, but as a strategic lever for sustaining the human and technical infrastructure on which critical open-source ecosystems depend.

The current literature underscores substantial contributions across multiple domains. AI-driven tools are being created and evaluated for automated maintenance activities, including intelligent bug identification, automated patch creation, and AI-facilitated code review, with the goal of reducing the strain on frequently overburdened maintainers and improving overall software quality. AI is also used to assess community health, providing metrics and insights into contributor engagement, potential attrition, and project vitality. In addition, AI-supported onboarding, task recommendation, and mentorship offer promising avenues for nurturing new contributors, while AI-based security analysis strengthens vulnerability detection and risk prediction. The growing use of AI-powered bots and agents for project management tasks, together with emerging work on \textbf{Green AI}, signals a shift toward more automated, data-driven, and energy-aware OSS practices.

At the same time, several critical gaps remain. There is little longitudinal evidence on how AI adoption reshapes the social dynamics, governance structures, and power relations within OSS communities, as most work still emphasizes short-term performance metrics rather than long-term sustainability outcomes. The moral and socio-technical implications of deploying AI in collaborative, volunteer-driven environments also remain underexplored, particularly around fairness, bias amplification, trust, and transparency in AI-mediated decisions such as triage, recommendation, or moderation. Moreover, while \textbf{Green AI} has emerged as an important theme, more research is needed to rigorously quantify the environmental footprint of AI solutions for OSS and to identify concrete practices that reduce energy consumption across the full model life cycle. Underlying all of these issues are data-related challenges: OSS datasets are heterogeneous, noisy, and context-dependent, which can hinder model robustness and generalizability if not carefully addressed.

The future research landscape in this interdisciplinary domain thus offers substantial opportunities. There is a clear need for more reliable and explicable AI systems tailored to OSS settings, emphasizing \emph{explainability}, \emph{controllability}, and \emph{human-in-the-loop} interaction so that maintainers and contributors can interrogate and adapt AI recommendations rather than simply accept them. Further work should explore methods to mitigate data scarcity and bias in OSS-based training corpora, and to understand how AI can foster inclusive, productive collaboration while respecting the established norms and values of open-source communities. For \textbf{Green AI}, promising directions include energy-efficient algorithms, carbon-aware experimentation workflows, and standardized sustainability metrics and reporting for AI-enhanced OSS projects.

Ultimately, it is imperative to develop integrated frameworks for the ethical and responsible use of AI within the OSS ecosystem, combining technical safeguards, governance mechanisms, and community-driven guidance. Artificial Intelligence possesses significant potential to reshape the sustainability of open-source software, but realizing this potential requires ongoing empirical study, careful design, and a sustained commitment to addressing the attendant risks. If approached thoughtfully, AI can function as a powerful facilitator rather than a disruptive force in cultivating a flourishing, resilient, and sustainable open-source ecosystem. In addition, future work would benefit from cross-disciplinary collaboration between software engineering, human-computer interaction, and sustainability research to build a more unified theoretical foundation. The development of benchmark datasets, shared evaluation frameworks, and replicable tooling could further accelerate scientific progress by reducing fragmentation across research efforts. Moreover, studying OSS ecosystems through longitudinal and comparative lenses may reveal patterns that short-term experiments cannot capture, particularly regarding contributor behavior, governance adaptations, and long-range project evolution. These directions collectively highlight that the intersection of AI and OSS sustainability remains wide open for rigorous and impactful scholarly inquiry.

\bibliographystyle{unsrt}  
\bibliography{references}

\end{document}